\documentclass[twocolumn,aps,floatfix,pra,showpacs]{revtex4}
\usepackage{amssymb,amsfonts,mathrsfs}
\usepackage{graphicx,bm}
\usepackage[T1]{fontenc}

\begin{document}

\title{Classical fields method for a relativistic interacting Bose gas}
\author{Emilia Witkowska$\,^1$, Pawe\l{} Zi\'n$\,^{2, \, 3}$ and Mariusz Gajda$\,^{1, \, 4}$}
\affiliation{
\mbox{$^1$ Institute of Physics, Polish Academy of Sciences, al.Lotnik\'ow $32/46$, 02-668 Warsaw, Poland}
\mbox{$^2$ Institute for Theoretical Physics, Warsaw University, ul. Ho\.za 69, 00-681 Warsaw, Poland}
\mbox{$^3$ Soltan Institute for Nuclear Studies, ul. Ho\.za 69, 00-681 Warsaw, Poland}
\mbox{$^4$ Faculty of Mathematics and Natural Sciences, College of Sciences, Cardinal's Stefan Wyszy\'nski University,}
\mbox{ul. Dewajtis $5$, 01-815 Warsaw, Poland}}
\date{\today}

\begin{abstract}
We formulate a classical fields method for description of relativistic interacting bosonic particles at nonzero temperatures. The method relays on the assumption that at low temperatures the Bose field can be described by a c-number function. We discuss a very important role of the cut-off momentum which divides the field into dominant classical part and a small quantum correction. We illustrate the method by studying a thermodynamics of relativistic Bose field which dynamics is governed by the Klein-Gordon equation with $\lambda \psi^4$ term responsible for interactions. 
\end{abstract}


\maketitle

\section{Introduction}
\label{sec: introduction}

Achievement of a Bose-Einstein condensation in dilute gases of alkali atoms initiated very extensive experimental and theoretical studies of these systems. In particular, a very important progress have been made in description of a Bose-Einstein condensate at nonzero temperatures. Various rigorous methods were developed 
to study thermal properties of Bose-Einstein condensates \cite{exact}. Unfortunately implementation of these methods is very difficult in realistic and physically interesting situations. Therefore other approximate approaches were considered. These are: the classical fields method based on the Gross-Pitaevskii equation (or projected Gross-Pitaevskii equation)
\cite{{cfm1},{cfm3}}, truncated Wigner method \cite{wigner}, stochastic Gross-Pitaevskii equation \cite{gardiner} or two fluid approach based on coupled Gross-Pitaevskii and kinetic equations \cite{zaremba}. In our opinion the methods based on the classical fields approximation are the most fruitful \cite{review}.

The classical fields method is based on the observation that a granular structure of the bosonic field is totally dominated by its wave character at temperatures below a condensation point. The bosonic field can be then described by a single complex function. This is the essential simplification.  It is justified only in a region of parameters were quantum fluctuations are much smaller than thermal ones. Evidently, classical fields method does not take into account spontaneous processes or the effect of quantum depletion. Advantage of the method is related to its great simplicity. It allows for studying not only thermal equilibrium properties but also nontrivial dynamics far from thermodynamic equilibrium. It can be relatively easy generalized to a system of many interacting Bose fields. Moreover, the method allows for a direct physical interpretation and deeper understanding of underlying physics. The classical fields method has been applied to description of a variety of thermal effects such as: a shift of the critical temperature due to interactions \cite{Davis}, dissipative dynamics of vortexes \cite{schmidt}, fragmentation of 1D condensates \cite{Kadio}, crystallization of vortex lattices \cite{lobo}, decay of doubly quantized vortex \cite{Gawryluk}, superfluity \cite{Zawitkowski-sup}, dynamics of spinor condensates \cite{spinor}, Kosterlitz-Thouless transition in 2D systems \cite{KBT}, a phase coherence \cite{Sinatra & Witkowska} or distillation \cite{Gorecka} of a Bose-Einstein condensate.

Encouraged by the success of the classical fields method we reformulate here this method in order to describe interacting relativistic Bose gas. We illustrate the method by studying thermodynamics of a relativistic Bose field describing system of relativistic massive particles and antiparticles whose dynamics is governed by the Klein-Gordon equation with $\lambda \psi^4$  term responsible for interactions. In a case of relativistic system we have to deal with additional degrees of freedom -- not only the field but also its momentum are dynamical variables. Moreover, the total number of particles or antiparticles is not conserved. Only their difference -- the total charge -- is preserved during the evolution \cite{{Bjorken},{BjorkenGreiner}}. All these differences must be taken into account while formulating the classical fields method for relativistic case.   

The system of degenerate relativistic gas of bosons was widely studied both in noninteracting \cite{{noninteracting1},{noninteracting2},{noninteracting3},{nieoddzialujacy RKBE},{noninteracting5},{noninteracting6}} as well as in interacting cases \cite{{Boisseau},{interacting},{Bernstein},{Haber},{kapusta},{Andersen}}. Bose Einstein condensation was predicted for such systems. The interacting case is much more interesting as some properties of the system can be drastically changed \cite{other}. In this case the condensation can be view as a spontaneous symmetry breaking in the $\lambda \psi^4$-theory \cite{kapusta}. The thermodynamic properties of the relativistic Bose-Einstein condensate were studied using the path-integral method \cite{{Haber},{kapusta}} and more recently using $2\Pi \, 1/N$ expansion \cite{Andersen} in the context of pion and kaon condensation. This method was used also for dynamical studies \cite{Equilibrium NLKGE}.
In fact a classical approximation, very similar to the one we want to introduce here,  has been used already in studies of  relaxation and thermalization in $\psi^4$ theory in 1+1 dimensions \cite{1+1}. It was found that the classical approach cannot describe this processes accurately as the system goes towards a classical not a quantum equilibrium. In our opinion this is because the problem of dynamical selection of the cut-off momentum was not treated with a proper care. The studies of the nonrelativistic systems show its extremely important role \cite{witkowska}. We will discuss this issue in detail in Sec. IV. 

In this paper we formulate the classical fields method for the relativistic $\psi^4$ theory in 3+1 dimensions. Unlike in previously studied 1+1 case \cite{1+1} the system exhibits a Bose-Einstein condensation and particle and antiparticle spectra are different in the region of small momenta. In order to extract a particle aspect of the field from the classical wave-like solution we first study the system analytically at zero temperature using Bogoliubov 
approximation \cite{Bogoliubov}, section III. This enables us to identify two types of quasiparticles, their energies and charges. This is crucial for further implementation. Then in section IV we formulate the classical fields method for the relativistic system and illustrate its application to a uniform system of relativistic, charged Bose field at a thermal equilibrium. We show how to determine important physical parameters, i.e. temperature, chemical potential or excitation energies. We stress a very important role of an appropriate choice of the cut-off momentum. Finally in section V, we show how the physical properties of the system vary with temperature. In particular we show that the system can be viewed as weakly interacting gas  of quasiparticles even very close to the critical temperature. Our particular results agree with the ones of \cite{{Bernstein},{Andersen},{kapusta}}.

\section{The model}
\label{sec: model}

We consider a relativistic quantum Bose field $\hat{\Psi} (\mathbf{r}, t)$ in a three dimensional cubic box of volume $V=L^3$ with periodic boundary conditions. We assume the standard `$\lambda {\psi}^4$' form of the interaction term. Then the Lagrangian of the system is $\hat{L} = \int d^3\mathbf{r} \, \hat{\mathcal{L}}$ with the Lagrangian density of the form:
\begin{equation}
\hat{\mathcal{L}} = \frac{1}{2 m} \left( \dot{\hat{\Psi}} ^{\dag} \dot{\hat{\Psi}} -
\nabla \hat{\Psi}^{\dag} \nabla \hat{\Psi}  - m^2 \hat{\Psi}^{\dag} \hat{\Psi} 
- \frac{\lambda}{2} \hat{\Psi}^{\dag} \hat{\Psi}^{\dag} \hat{\Psi} \hat{\Psi} \right) \, ,
\label{eq: lagrangian density}
\end{equation}
where $\nabla$ is the three dimensional gradient, dot denotes time derivative and $m$ is a mass ($m^2>0$). Our units are described in \cite{units}. Momentum conjugated to the field is:
\begin{equation}
\hat{\Pi}(\mathbf{r}, t) = \frac{1}{2m} \dot{\hat{\Psi}}^{\dag} (\mathbf{r}, t) \, .
\label{eq: moments}
\end{equation}
The field and the momentum satisfy bosonic commutation relation:
\begin{equation}
\left[ \hat{\Psi}(\mathbf{r}, t), \hat{\Pi} (\mathbf{r'}, t) \right] = 2 i m \, 
\delta^3 (\mathbf{r} - \mathbf{r'})
\label{eq: commutations 2}  \, ,
\end{equation}
while other commutators are equal to zero \cite{Bjorken}.

Corresponding Hamiltonian of the system is $\hat{H} = \int d^3\mathbf{r} \, \hat{\mathcal{H}}$ with the Hamiltonian density:
\begin{eqnarray}
\hat{\mathcal{H}} &=& \frac{1}{2 m} \left( (2 m)^2 \hat{\Pi} \hat{\Pi}^{\dag} + 
\nabla \hat{\Psi}^{\dag} \nabla \hat{\Psi}  \right. \nonumber \\
&+& \left. m^2 \hat{\Psi}^{\dag} \hat{\Psi}
+ \frac{\lambda}{2} \hat{\Psi}^{\dag} \hat{\Psi}^{\dag} \hat{\Psi} \hat{\Psi} \right) \, .
\label{eq: hamiltonian}
\end{eqnarray}
The field evolves in time according to the nonlinear Klein-Gordon equation (NLKGE):
\begin{equation}
\ddot{\hat{\Psi}} = \nabla^2 \hat{\Psi} - m^2 \hat{\Psi} 
- \lambda \hat{\Psi}^{\dag} \hat{\Psi} \hat{\Psi}\, .
\label{eq:equation of motion}
\end{equation}
The Hamiltonian commutes with the charge operator $\hat{Q} = \int d^3\mathbf{r} \, \hat{\mathcal{Q}}$, where the charge density $\hat{\mathcal{Q}}$ is defined as:  
\begin{equation}
\hat{\mathcal{Q}} = \frac{i}{2m} \left( \hat{\Psi}^{\dag} \dot{\hat{\Psi}} 
- \hat{\Psi} \dot{\hat{\Psi}}^{\dag} \right) \, .
\label{eq: charge}
\end{equation}
Therefore, the charge of the system is preserved in the evolution. In this paper we concentrate on the situation where the total charge is different than zero, hence without any loss of generality we assume that the charge is positive $\hat{Q}>0$.

\section{Bogoliubov transformation}
\label{sec: bogoliubov}

We assume that a Bose-Einstein condensate is present in the system and the lowest energy mode becomes macroscopically occupied.
In such a case annihilation and creation operators of this mode can be substituted by c-numbers. 
This is the famous Bogoliubov approximation \cite{Bogoliubov}.
Within this assumption the field operator $\hat{\Psi} (\mathbf{r}, t)$ takes the form:
\begin{equation}
\hat{\Psi} (\mathbf{r}, t) = \left( \phi_0 \, 
+ \sum_{\mathbf{k}} 
\hat{A}(\mathbf{k},t) e^{i \mathbf{k} \mathbf{r}} \right)\, e^{-i \mu t} \, ,
\label{eq: anzaltz}
\end{equation}
where the classical field $\phi_0$ represents a condensate,
$\mu$ is a positive energy and $\hat{A}(\mathbf{k},t)$ are Fourier components (with wave-vector ${\bf k}$)
of the field operator $\hat{\Psi} (\mathbf{r}, t)$.

\subsection{Excitation energies}

The next step of the Bogoliubov method relies on the condition that {\it almost all} 
particles are in $k=0$ mode. This makes us to neglect all terms in the Hamiltonian 
that describe the interaction of particles outside the condensate with themselves, 
keeping only terms responsible for interactions of excited particles with the condensate.
Technically it means that only linear terms in $\hat{A}(\mathbf{k},t)$
are kept in the nonlinear Klein-Gordon equation.
By comparing terms with the same time dependence we get two equations.
The first relates a value of $\mu$ to the condensate field:
\begin{equation}
\mu^2 = m^2 + \lambda |\phi_0|^2 \, ,
\label{eq: chemical potential}
\end{equation}
while the second determines dynamics of amplitudes $\hat{A}(\mathbf{k},t)$:
\begin{equation}
\ddot{\hat{A}}(\mathbf{k} ,t) = 2 i \mu \dot{\hat{A}}(\mathbf{k} ,t) - L_\mathbf{k} \hat{A}(\mathbf{k},t) 
- \lambda \phi_0^2 \hat{A}^{\dag}(-\mathbf{k},t) \, ,
\label{eq: equation for A}
\end{equation}
where $L_\mathbf{k} = \mathbf{k}^2 + \lambda |\phi_0|^2$. The conjugated amplitude 
$\hat{A}^{\dag}(-\mathbf{k},t)$ satisfies analogical equation. 
Let us note that the chemical potential is larger than the bare mass so the system is Bose condensed \cite{kapusta} consistently with our assumption (\ref{eq: anzaltz}).

These two second order equations have nontrivial solutions if the corresponding determinant vanishes:
\begin{equation}
(\omega_\mathbf{k}^2 - 2 \mu \omega_\mathbf{k} - L_\mathbf{k})
(\omega_\mathbf{k}^2 + 2 \mu \omega_\mathbf{k} - L_\mathbf{k}) = (\lambda |\phi_0|^2)^2 \, .
\label{eq: eq on omega}
\end{equation}
The above equation has two solutions (for every mode $\mathbf{k}$):
\begin{eqnarray}
\omega_{1} ^2  &=& \mathbf{k}^2 + 2 \mu^2 + \lambda |\phi_0|^2 - \nonumber \\
& - &\sqrt{4\mathbf{k}^2 \mu^2 + (2 \mu^2 + \lambda |\phi_0|^2)^2} \, ,
\label{eq: omega1}
\end{eqnarray}
and
\begin{eqnarray}
\omega_{2}  ^2 &=& \mathbf{k}^2 + 2 \mu^2 + \lambda |\phi_0|^2 + \nonumber \\
&+&\sqrt{4\mathbf{k}^2 \mu^2 + (2 \mu^2 + \lambda |\phi_0|^2)^2} \, .
\label{eq: omega2}
\end{eqnarray}
In a long wave-length limit ($\mathbf{k}\to 0$) the frequency $\omega_1(\mathbf{k})$ is:
\begin{equation}
\omega_{1} (\mathbf{k}) \cong \mathbf{k} \mathrm{v}_s \, ,
\label{eq: shortK}
\end{equation}
where $\mathrm{v}_s $ is a sound velocity:
\begin{equation}
\mathrm{v}_s = \Big( \frac{\lambda |\phi_0|^2}{2 \mu^2 + \lambda |\phi_0|^2} \Big)^{1/2}\, .
\end{equation}
Evidently these are phonon-like excitations typical for massless  particles. 
The linear character of the spectrum at small momenta is responsible for a superfluity of the system. 
In the long wave-length limit the second solution (\ref{eq: omega2}) has the form:
\begin{equation}
\omega_{2} (\mathbf{k}) \cong 
\left(\mathbf{k}^2 \frac{\mu^2}{(2 \mu^2 + \lambda |\phi_0|^2)^{2}} + 1\right)
\sqrt{2(2 \mu^2 + \lambda |\phi_0|^2)} \, .
\end{equation}
This dispersion relation is characteristic for massive particles.
Let us notice that at zero momentum $\mathbf{k} = 0$ we have: $\omega_{1} (0) = 0$ and 
$\omega_{2} (0) = \sqrt{2(2 \mu^2 + \lambda |\phi_0|^2)}$.
In a short wavelength limit ($\mathbf{k} \to \infty$) both branches of the dispersion relation are typical for relativistic noninteracting massive bosons:
\begin{eqnarray}
\omega_{1}({\bf k}) &=& \sqrt{{\bf k}^2 + m_{eff}^2} - \mu \, ,\\
\omega_{2}({\bf k}) &=& \sqrt{{\bf k}^2 + m_{eff}^2} + \mu \, ,
\end{eqnarray}
where the effective mass is $m_{eff}^2= \mu^2 + \lambda |\phi_0|^2$.
The only role of interactions is to modify a bare mass of particles.

The same spectrum of energies $\omega_{1}(\mathbf{k})$ and $\omega_{2}(\mathbf{k})$ has been found in
\cite{Bernstein} where the path-integral method were used and in \cite{Andersen} within $2 \Pi \, 1/N$ expansion.

\subsection{Eigenmodes}
\label{sec: amplitudes}

The results of the previous subsection allow us to write the 
operator $\hat{A}(\mathbf{k},t)$ in the form:
\begin{equation}
\hat{A}(\mathbf{k},t) = 
\hat{a}_\mathbf{k} e^{-i \omega_1 t} + \hat{b}_\mathbf{k} e^{i \omega_1 t} 
+ \hat{c}_\mathbf{k} e^{-i \omega_2 t} + \hat{d}_\mathbf{k} e^{i \omega_2 t} \, ,
\label{eq: A before}
\end{equation}
where $\hat{a}_\mathbf{k}, \,\hat{b}_\mathbf{k}, \,\hat{c}_\mathbf{k}, \, \hat{d}_\mathbf{k}$ are some operators
and $\omega_1$, $\omega_2$ are positive.
Let us notice that terms proportional to $\hat{a}_0$ and $\hat{b}_0$ do not depend on time. 
Such constant terms are already included in $\phi_0$.
Therefore we have:
\begin{eqnarray}
\hat{a}_0&=&0\, ,
\label{eq: a0 jest 0}\\
\hat{b}_0&=&0\, .
\end{eqnarray}
If $\hat{A}(\mathbf{k},t)$ satisfies equation (\ref{eq: equation for A})
then the operators $\hat{b}_\mathbf{k}$ and $\hat{c}_\mathbf{k}$ are: 
\begin{eqnarray}
\hat{b}_\mathbf{k} &=& \gamma_1(\mathbf{k}) \, \hat{a}_{-\mathbf{k}}^{\dag}, \\
\hat{c}_\mathbf{k} &=& \gamma_2(\mathbf{k}) \, \hat{d}_{-\mathbf{k}}^{\dag},
\end{eqnarray}
where
\begin{eqnarray}
\gamma_1(\mathbf{k}) &=& \frac{\lambda \phi_0^2}{w_1^2 - 2 \mu \omega_1 - L_\mathbf{k}} \, , 
\label{eq: gamma1}\\
\gamma_2(\mathbf{k}) &=& \frac{\lambda \phi_0^2}{w_2^2 + 2 \mu \omega_2 - L_\mathbf{k}} \, .
\label{eq: gamma2}
\end{eqnarray}
Thus we have:
\begin{equation}
\hat{A}(\mathbf{k},t) = \hat{a}_\mathbf{k} e^{-i w_1 t} + \gamma_1 \hat{a}_{-\mathbf{k}}^{\dag} e^{i w_1 t}  
+ \hat{d}_\mathbf{k} e^{i w_2 t} + \gamma_2 \hat{d}_{-\mathbf{k}}^{\dag} e^{-i w_2 t} \, .
\label{eq: A}
\end{equation}
Commutation relations (\ref{eq: commutations 2}) impose additional 
constrains on operators $\hat{a}_\mathbf{k}, \, \hat{d}_\mathbf{k}$.
To explore this fact we introduce scaled operators $\hat{\alpha}_\mathbf{k}$ and $\hat{\beta}_\mathbf{k}$:
\begin{eqnarray}
\hat{a}_\mathbf{k} & = & f_\mathbf{k} \,  \hat{\alpha}_\mathbf{k} \, , \nonumber \\
\hat{d}_\mathbf{k} & = & g_\mathbf{k} \, \hat{\beta}_{-\mathbf{k}}^{\dag} \, ,
\label{eq: 26}
\end{eqnarray}
where, at this stage, $f_\mathbf{k}$ and $g_\mathbf{k}$ are some arbitrary functions. 
Consistently with (\ref{eq: a0 jest 0}) $f_{0}$ is equal to zero:
\begin{equation}
f_{0}=0 \, .
\label{eq: f0 jest 0}
\end{equation}
According to (\ref{eq: commutations 2}) we assume bosonic commutation relations
for the scaled operators:
\begin{eqnarray}
\label{eq: alphabeta1}
[\alpha_\mathbf{k}, \alpha_\mathbf{k'}^{\dag}] &=& \delta_{\mathbf{k} \mathbf{k'}} \, ,\\
\label{eq: alphabeta2}
[\beta_\mathbf{k}, \beta_\mathbf{k'}^{\dag}] &=& \delta_{\mathbf{k} \mathbf{k'}} \, ,
\end{eqnarray}
what allows to determine functions $f_\mathbf{k}$, $g_\mathbf{k}$. 
In the momentum representation the field operator takes the form:
\begin{equation}
\hat{A}(0, t) e^{+i \mu t} = g_0 \left( \hat{\beta}_0^{\dag} (t) + \gamma_2(0) \hat{\beta}_0 (t) \right) \, ,
\label{eq: filed 0}
\end{equation}
for ${\bf k} = 0$, and $\hat{A}(\mathbf{k},t)=\hat{A}_{1}(\mathbf{k},t) + \hat{A} ^\dag_{2}(-\mathbf{k},t)$ with:
\begin{eqnarray}
\hat{A}_{1}(\mathbf{k},t) e^{+ i \mu t}&=& f_\mathbf{k} \left( \hat{\alpha}_\mathbf{k} (t) + 
\gamma_1(\mathbf{k}) \hat{\alpha}_\mathbf{-k}^{\dag} (t) \right) \, , \label{eq: filed P} \\
\hat{A} ^\dag_{2}(-\mathbf{k},t) e^{+ i \mu t} &=& g_\mathbf{k} \left( \hat{\beta}_\mathbf{-k}^{\dag}(t) 
+ \gamma_2(\mathbf{k}) \hat{\beta}_\mathbf{k}(t) \right) \, , 
\label{eq: filed}
\end{eqnarray}
for ${\bf k} \ne 0$. In the above equations we used following notation 
$\hat{\alpha}_\mathbf{k} (t)= \hat{\alpha}_\mathbf{k} e^{- i \omega_1({\bf k}) t}$ and
$\hat{\beta}_\mathbf{-k}^\dag (t)= \hat{\beta}_\mathbf{-k}^\dag e^{+ i \omega_2({\bf k}) t}$.
Functions $ f_\mathbf{k}$ and $ g_\mathbf{k}$ are:
\begin{eqnarray}
|f_\mathbf{k}|^2 &=& \frac{z_\mathbf{k}}{1 - |\gamma_1(\mathbf{k})|^2} \, ,
\label{eq: fk}\\
|g_\mathbf{k}|^2 &=& \frac{z_\mathbf{k}}{1 - |\gamma_2(\mathbf{k})|^2} \, ,
\label{eq: gk}
\end{eqnarray}
where:
\begin{equation}
z_\mathbf{k} = \frac{4 m \mu}{\omega_2^2({\bf k}) - \omega_1^2({\bf k})} \, .
\label{eq: zk}
\end{equation}
Functions $f_{\bf k}$ and $g_{\bf k}$ are known up to some arbitrary phase factors.
In what follows we set the phases to zero.

Equations (\ref{eq: filed P}) and (\ref{eq: filed}) define the relativistic Bogoliubov transformation.
In the noninteracting limit ($\lambda \to 0$) operators 
$\hat{\alpha}_{{\bf k}}^\dag$ and $\hat{\beta}_{{\bf k}}^\dag$ create a bare (noninteracting)
particle and antiparticle respectively. 
In the presence of interactions the system consists of two types of quasiparticles.
The operator  $\hat{\alpha}_{{\bf k}}^\dag$ creates quasiparticles with energy $\omega_1({\bf k})$ and the operator $\hat{\beta}_{{\bf k}}^\dag$ creates quasiparticles with energy $\omega_2({\bf k})$.
The first type of excitations are quasiparticles which in the small momentum limit
are characterized by the phonon-like dispersion relation.
Excitations of the second type are typical relativistic massive particles. 
These quasiparticles are mixtures of bare particles and bare antiparticles. 
Coefficients $\gamma_1$ and $\gamma_2$ determine an amount of this mixing.

\subsection{Hamiltonian and charge}
\label{sec: Hamiltonian and charge}

In the previous sections we have analyzed the equation of motion of the field operator in the 
Bogoliubov approximation. Here we want to approximate the Hamiltonian of the system
using arguments of the Bogoliubov method.
Consistently we keep in the Hamiltonian terms proportional to the nonzero value of the field 
and terms proportional to amplitudes $\hat{A}_\mathbf{k}$ up to the second order.
However, this approximated Hamiltonian does not conserve the field's charge $Q$. 
In order to control a mean value of the charge we introduce the operator:
\begin{equation}
\hat{F} = \hat{H} - \tilde \mu \hat{Q} \,  ,
\label{eq: F}
\end{equation}
where $\tilde \mu$ is a chemical potential and plays a role of the Lagrange multiplier.
The Bogoliubov transformation (\ref{eq: filed 0})-(\ref{eq: filed}) brings the Hamiltonian
to the diagonal form provided that the Lagrange multiplier $\tilde \mu$ in (\ref{eq: F}) 
is equal to the frequency $\mu$ of the zero momentum component  of the field (\ref{eq: anzaltz}):
\begin{equation}
\hat{F} = \sum_{\mathbf{k}} \left[ \omega_1(\mathbf{k}) \left( \hat{\alpha}_{\mathbf{k}}^{\dag} \hat{\alpha}_{\mathbf{k}}
+ {1 \over 2}\right) + \omega_2 (\mathbf{k}) \left( \hat{\beta}_{\mathbf{k}}^{\dag} \hat{\beta}_{\mathbf{k}} + {1 \over 2}\right)\right] .
\label{diag}
\end{equation}
As we see $\mu$ is a chemical potential. 
In (\ref{diag}) we skipped the irrelevant constant term $ -{\lambda}/(4m) |\phi_0|^4$.
The diagonal form of $\hat{F}$ shows again that we deal with the system of noninteracting quasiparticles.
A number of the first quasiparticles with momentum ${\bf k}$ is obviously:
\begin{equation}
\hat{N}_{1}(\bf k) =\hat{\alpha}_{\bf k}^{\dag} \hat{\alpha}_{\bf k} \, ,
\end{equation}
while the number of the second type of quasiparticles has the form:
\begin{equation}
\hat{N}_{2}(\bf k) =\hat{\beta}_{\bf k}^{\dag} \hat{\beta}_{\bf k} \, .
\end{equation}

The same approximations bring the total charge to the form:
\begin{equation}
\hat{Q} \simeq \frac{\mu}{m}|\phi_0|^2 +\mathcal{C} + \sum_{{\bf k}} \left( q_{1}({\bf k}) \hat{N}_{1}({\bf k})
+ q_{2}({\bf k}) \hat{N}_{2}({\bf k}) \right) \, ,
\label{eq: charge final}
\end{equation}
where
\begin{eqnarray}
q_{1}({\bf k})&=&\frac{2\mu}{\omega_1} \, \frac{L_{\bf k} + \omega_1^2}{\omega_2^2-\omega_1^2},
\label{eq: qP} \\
q_{2}({\bf k})&=&-\frac{2\mu}{\omega_2} \, \frac{L_{\bf k} + \omega_2^2}{\omega_2^2-\omega_1^2}\, ,
\label{eq: qA}
\end{eqnarray}
and
\begin{eqnarray}
\mathcal{C}&=&\sum_{k}\left\{|f_{\bf k}|^2 |\gamma_1|^2 \left(1 + \frac{\omega_1({\bf k})}{\mu}\right) +\right.\nonumber \\
{}&+&\left.|g_{\bf k}|^2 |\gamma_2|^2 \left(1 - \frac{\omega_2({\bf k})}{\mu}\right)\right\} \, .
\label{eq: C 11}
\end{eqnarray}
In obtaining equation (\ref{eq: charge final}) rapidly oscillating terms have been neglected. They will not contribute to a value of the charge averaged over time. Quasiparticles of both types are carrying a definite charge (in terms of time averaged quantities). It turns out that $q_{1}(k)$ is positive while $q_{2}(k)$ is negative. In the limit of large $k$ the charge of the first quasiparticles tends to $q_{1} \to 1$ and the charge of the second quasiparticles tends to $q_{2} \to -1$. Assuming thermal Bose-Einstein distribution of both kind of quasiparticles it is easy to check that the sum
in (\ref{eq: charge final}) converges. The quantity $\mathcal{C}$ in (\ref{eq: charge final}) is a quantum depletion.
It gives the amount of the charge missing from $k=0$ mode in the ground state of the many-body system. The quantum depletion is proportional to $\lambda |\phi_0|^2/m^2$ which is a small parameter in the case studied here \cite{quantum depletion value}. Thus at zero temperature the whole charge is accumulated in ${\bf k}=0$ mode and a zero-momentum field amplitude is: 
\begin{equation}
|\phi_0|^2  = \langle \hat{Q} \rangle \frac{m}{\mu}  \, .
\label{eq: Q0}
\end{equation}
Equations (\ref{eq: Q0}) and (\ref{eq: chemical potential})
determine values of the chemical potential $\mu$ and the condensate field intensity $|\phi_0|^2$.
The excitation spectrum and the time evolution of the field operator are determined uniquely.

\section{Formulation of the classical fields method}
\label{sec: simulations}

\subsection{General considerations}

While formulating the classical fields method for relativistic bosons we follow the procedure developed for nonrelativistic case \cite{review}. We assume periodic boundary conditions and then the natural modes of the system are plane waves.
Expansion of the field operator $\hat{\Psi}$ and its momentum $\hat{\Pi}$ in the basis of plane waves gives:
\begin{eqnarray}
\hat{\Psi} (\mathbf{r}, t) = \sum_{{\bf k}} \hat{\Psi} (\mathbf{k}, t) e^{i {\bf k r}}\, , \\
\hat{\Pi} (\mathbf{r}, t) = \sum_{{\bf k}} \hat{\Pi} (\mathbf{k}, t) e^{i {\bf k r}}\, ,
\end{eqnarray}
where ${\bf k}$ is restricted to the first Brillouin zone and takes values ${\bf k}=(2 \pi/L) {\bf n}$ with $n_i=0, \pm 1, \pm 2 \dots$ ($i=x,y,z$), and $L$ is a length of the cubic box. If a population of a given mode is greater than its quantum fluctuations it is legitimate to replace corresponding operators by c-numbers. In practice even the mode whose occupation is of the order of one can be treated as classical. At zero temperature it is only the zero momentum mode ${\bf k}=0$ which fulfills this condition. The field operator can be divided into a large 'classical' part and a small 'quantum' corrections. This is the essence of the famous Bogoliubov approximation \cite{Bogoliubov}. 
Evidently, at higher temperatures many modes are macroscopically populated. Hence, the 'classical' part of the field includes not only the zero momentum component but also components of larger momenta up to some value $k_{max}$. Occupations of remaining modes are set to zero in the lowest order of approximation. Consistently, the field operator $\hat{\Psi}$ and corresponding momentum $\hat{\Pi}$ are approximated by:
\begin{eqnarray}
\hat{\Psi} (\mathbf{r}, t) \to \Psi (\mathbf{r}, t) = \sum_{|{\bf k}|\le k_{max}} \Psi (\mathbf{k}, t) e^{i {\bf k r}}\,  , \\
\hat{\Pi} (\mathbf{r}, t) \to \Pi (\mathbf{r}, t) = \sum_{|{\bf k}|\le k_{max}} \Pi (\mathbf{k}, t) e^{i {\bf k r}}\,  ,
\end{eqnarray}
where $\Psi$ and $\Pi$ are c-number functions and $k_{max}$ is a cut-off momentum. 

The complex function $\Psi({\bf r},t)$ satisfies the same nonlinear Klein-Gordon 
equation (\ref{eq:equation of motion}) as the field operator. 
It is convenient to rescale the field in such a way
that its total charge is equal to one, $\Psi \to \sqrt{Q} \Psi$. After such scaling the equation
satisfied by the classical field is:
\begin{equation}
\ddot{\Psi}({\bf r},t)=\nabla^2 \Psi({\bf r},t)-m^2 \Psi({\bf r},t)
- \lambda Q |\Psi({\bf r},t)|^2 \Psi({\bf r},t)\, ,
\label{re: nlkge skalowane}
\end{equation}
and the field $\Psi({\bf r},t)$ has the unit charge. Let us notice that the charge $Q$ and the interaction strength
$\lambda$ enter the dynamical equation through the product $Q\lambda$ only.
However, physical properties of the system at finite temperature depend separately on $Q$ and $\lambda$. 
These very important issue will be discussed later.

A steady state evolution of the field whose dynamics is governed by Eq.(\ref{re: nlkge skalowane}) corresponds to a single realization of the relativistic system at a thermal equilibrium. Based on the ergodic hypothesis, time averages of physical observables with respect to the field fluctuating around its mean value correspond to the microcannonical description.  The only problem is to find a steady state solution corresponding to a given energy (temperature) and charge. Fortunately, due to the nonlinear term, all solutions of Eq.(\ref{re: nlkge skalowane}) for a given energy and charge tend to the same equilibrium state independently on the particular choice of initial conditions. We checked this observation by solving numerically the dynamical equations for different random choice of the initial field $\Psi({\bf r},0)$ and momentum $\Pi({\bf r},0)$.  

Let us note that because a number of classical modes is restricted by the cut-off momentum the equation (\ref{re: nlkge skalowane}) is defined on a well specified finite spatial grid $N_{grid}=\mathcal{N}^3$ and the cut-off momentum is $k_{max}=\sqrt{3} \pi \mathcal{N}/L$. Therefore, while solving Eq.(\ref{re: nlkge skalowane}) we have to choose a number of classical modes first. This number must depend on the temperature (or the energy) of the system. However, the value of cut-off can be precisely known {\it a posteriori} when an equilibrium state is generated numerically and thermal occupations are determined. Therefore, the choice of the cut-off momentum is based on a smart guess which is verified when the full analysis of solution is done. 

The quantitative criterion for an optimal choice of the cut-off momentum is directly related to the fundamental assumption of the classical fields method -- a macroscopic occupation of modes. In order to determine occupation of modes we have to be able to assign a unique value of a number of particles (which does not have to be an integer) to every given amplitude of the $\bf k$-momentum mode, $|\Psi({\bf k})|^2$. In our approach the link between a field amplitude and a relative occupation of ${\bf k}$-momentum mode follows directly from the Bogoliubov approximation. Therefore, the standard zero temperature Bogoliubov approach becomes an inherent part of the classical fields method. Various studies of the optimal choice of the cut-off momentum in the nonrelativistic theory \cite{zawitkowski, witkowska, review} indicate that a `field intensity' (i.e. modulus square) of the highest momentum wave defines the unit which corresponds to a nonrelativistic single particle. This allows for determination of all modes occupations provided that their intensities are known. Moreover, as shown in \cite{witkowska} this procedure allows for `quantization' of the classical field and ensures that thermal particle distribution resembles the Bose statistics. Such quantization procedure we generalize to the relativistic case.         

Let us note that the dynamics can lead to states of momenta larger than $k_{max}$ because of the nonlinear term.
In some implementations of the method a projection operator in space coordinates is introduced. This operator is applied at every time step to ensure that only projected part of the field is considered \cite{Davis}. In our implementation, 
in order to restrict dynamics to classical modes, $|{\bf k}|<k_{max}$, all momenta are restricted to the first Brillouin zone automatically due to periodic boundary conditions and usage of the Fast Fourier Transform method for time propagation.

All these general considerations will be elaborated in details in the next subsection where we study a thermal equilibrium of the relativistic massive charged Bose field.

\subsection{Numerical implementation}

Starting from a random initial condition we evolve the classical field according to Eq.(\ref{re: nlkge skalowane}). 
The first important observation is that after some time the field reaches a steady state: 
amplitudes of different momenta modes fluctuate around some mean values.
Then we look into details of the equilibrium state by analyzing the spectrum of amplitudes:
\begin{equation}
G({\bf k}, \omega) = |\Psi({\bf k}, \omega)|^2 \, ,
\label{eq: green fun}
\end{equation}
where
\begin{equation}
\Psi({\bf k}, \omega) = \sum_{{\bf r}}  e^{i {\bf k\, r}} \sum_{t}  e^{-i \omega t} \Psi({\bf x}, t)  \, .
\end{equation}
Notice, the spectrum $G({\bf k}, \omega)$ is proportional to the Green function \cite{Green comment}.

\begin{figure}[]
\centerline{\includegraphics[width=0.38\textwidth, height=0.38\textheight, keepaspectratio,angle=-90]{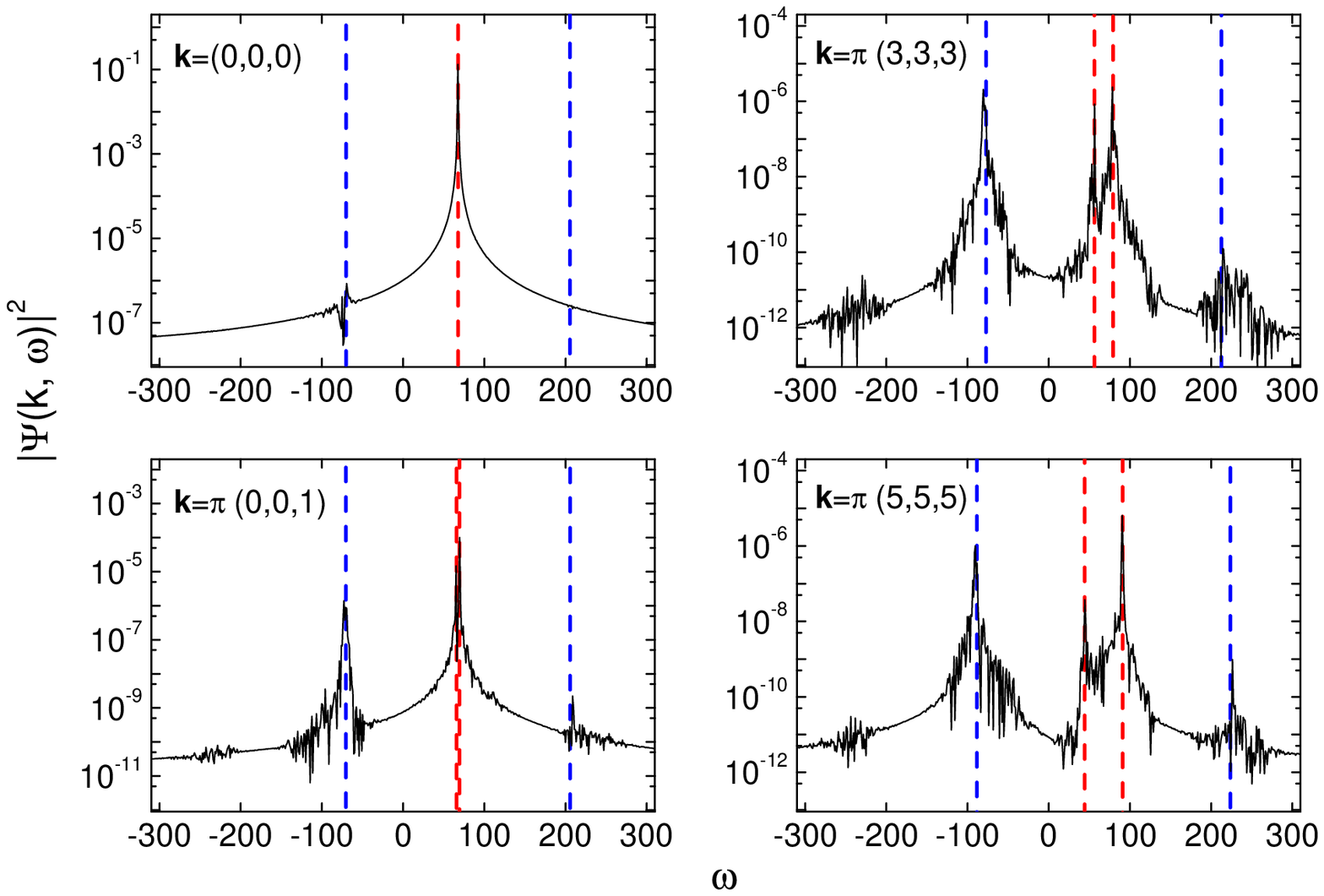}}
\caption{(Color online) Spectrum of amplitudes. The black solid lines refer to
the numerical simulations, dashed lines refer to the peaks' positions
given by the Bogoliubov-Popov like formulas (see text for explanation). For the zero momentum mode the position of the highest peak is $\mu(T)$
while its amplitude is $|\phi_0(T)|^2$. The left peak corresponds to $\omega^{-}_2(0)$ 
(left dashed blue line), and the right peak to
$\omega^{+}_2(0)$ (right dashed blue line). Notation for the others peaks ($k \ne 0$)
is the following: $\omega^{-}_1({\bf k})$ -- red left line, $\omega^{+}_1({\bf k})$ -- red right line,
and $\omega^{-}_2({\bf k})$ -- blue left line, and $\omega^{+}_2({\bf k})$ -- blue right line.
Parameters are: mass $m=60$, $\lambda Q=900$, energy $E=75$ and 
the number of grid points $\mathcal{N}^3=16^3$.}
\label{fig: spectrum of amplitudes1}
\end{figure}

In Fig. \ref{fig: spectrum of amplitudes1} we plot spectrum of amplitudes (\ref{eq: green fun})
for several values of momentum $\bf k$. In order to show `fine structure details' we use the logarithmic scale. 
Positions of characteristic peaks (`singularities') marked by the vertical dashed lines correspond to  
excitation energies of quasiparticles. A width of each peak results from
quasiparticles interactions and is related to a finite lifetime of a quasiparticle in a given mode.
For the zero momentum mode ($k=0$) one can observe three distinct peaks in the spectrum. 
The amplitude of the central one dominates.
The averaged position of the highest peak we denote by $\mu(T)$ (red dashed line),
and will refer to it as to a chemical potential. The amplitude of this peak can be associated with
$|\phi_0(T)|^2$. Positions of the other two peaks are: $\omega^{+}_2(0)=\mu(T) + \omega_2(0)$ 
(right dashed line) and $\omega^{-}_2(0)=\mu(T) - \omega_2(0)$ (left dashed line).
The spectrum of ${\bf k} \ne 0$ mode is composed of four peaks.
Positions of these peaks are symmetric with respect to $\mu(T)$.
In general characteristic frequencies visible in the spectrum can be divided into two branches: 1) $\omega^{\pm}_{1}({\bf k})=\mu(T) \pm \omega_{1}({\bf k})$ , and 2) $\omega^{\pm}_{2}({\bf k})=\mu(T) \pm \omega_{2}({\bf k})$.

\begin{figure}[]
\centerline{\includegraphics[width=0.28\textwidth, height=0.28\textheight, keepaspectratio,angle=-90]{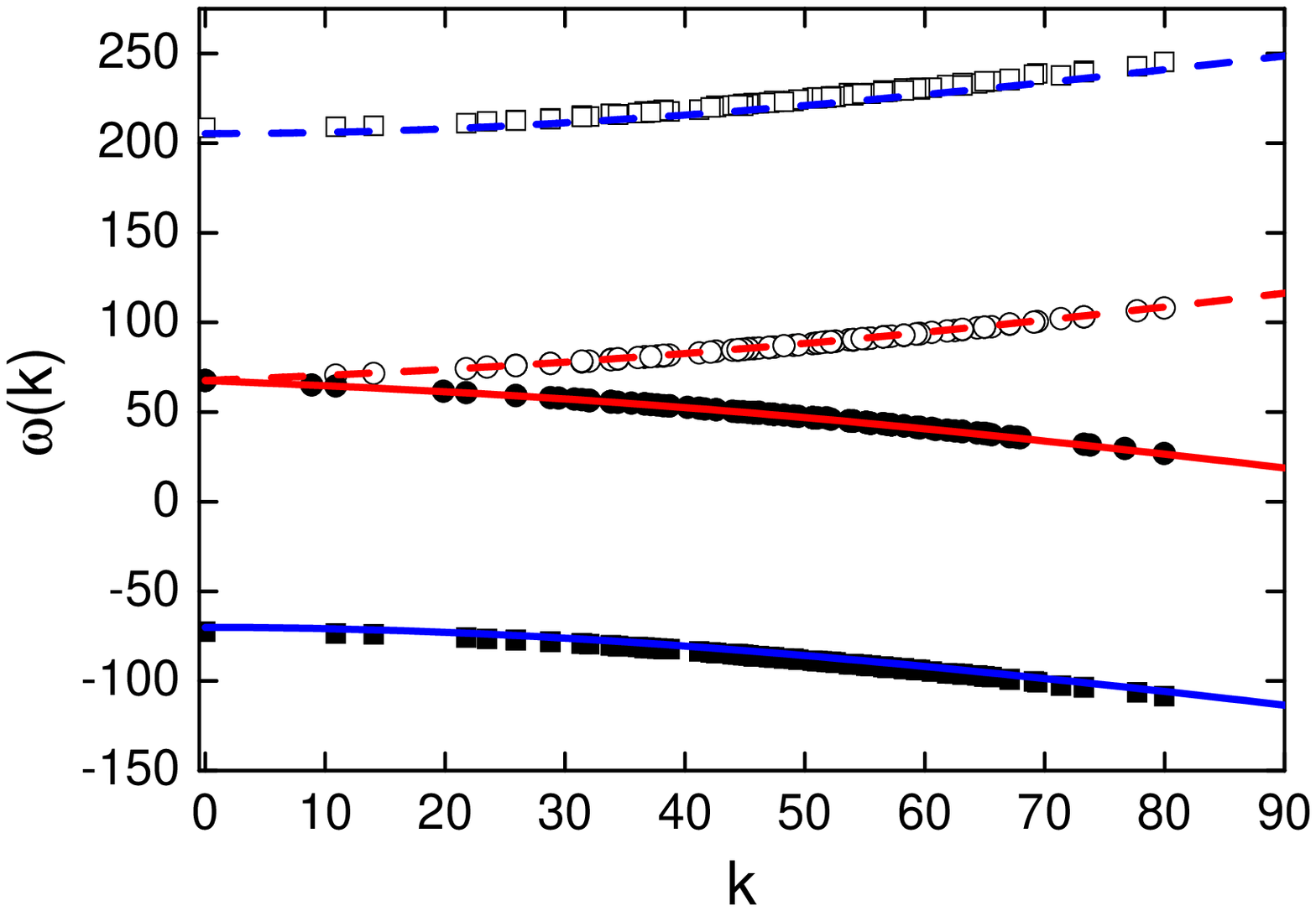}}
\caption{(Color online) Excitation energy of quasiparticles.
Points refer to peaks' positions calculated numerically according to (\ref{eq: omega cfs}).
Lines refer to the Bogoliubov-Popov like formulas (see text for explanation).
$(i)$ empty circles (upper dashed red line) $\omega^{+}_1({\bf k})$, 
$(ii)$ full circles (lower red line) $\omega^{-}_1({\bf k})$,
$(iii)$ empty box (upper dashed blue line) $\omega^{+}_2({\bf k})$ 
and $(iv)$ full box (lower blue line) $\omega^{-}_2({\bf k})$.
Parameters are the same as in figure \ref{fig: spectrum of amplitudes1}.}
\label{fig: Energy spectrum}
\end{figure}

Central frequencies of peaks we calculated as follows:
\begin{equation}
\omega({\bf k}) = \frac{\sum_{\omega \in S} \omega \, |\Psi({\bf k}, \omega)|^2}
{\sum_{\omega \in S} |\Psi({\bf k}, \omega)|^2} \, ,
\label{eq: omega cfs}
\end{equation}
where $S$ is a frequency range defined by the width of the peak. In Fig. \ref{fig: Energy spectrum} we show results of numerical simulations (marked by circles and squares) and formulas (\ref{eq: omega1}) and (\ref{eq: omega2}) with $\mu(T)$ and $|\phi_0(T)|^2$ being determined numerically from the evolution of the ${\bf k}=0$ mode (marked by lines).
In the nonrelativistic theory this modification of the zero temperature spectrum is known as the Bogoliubov-Popov relation (BP) \cite{BP}. We notice perfect agreement between numerical results and the BP-like formulas. Let us notice that the first type of excitations $\omega_{1}({\bf k})$ (red lines) is characterized by a linear dispersion relation for small ${|\bf k|}$. Excitations of the second type, $\omega_{2}({\bf k})$, (blue lines) have particle like quadratic dispersion in the small ${|\bf k|}$ region.

If we neglect peak's width then the stationary state of the field in the momentum space can be approximated by:
\begin{eqnarray}
\Psi({\bf k}, t) e^{+i\mu t} &=&
\mathcal{A}^{+}_1({\bf k}) e^{-i \omega_1({\bf k}) t}
+\mathcal{A}^{-}_1({\bf k}) e^{i \omega_1({\bf k}) t} \nonumber \\
&+&\mathcal{A}^{+}_2({\bf k}) e^{-i \omega_2({\bf k}) t}
+ \mathcal{A}^{-}_2({\bf k}) e^{i \omega_2({\bf k}) t} \, ,
\label{eq: przyblizenie num}
\end{eqnarray}
with amplitudes $\mathcal{A}^{\pm}_{1,2}({\bf k})$ calculated according to:
\begin{equation}
|\mathcal{A}^{\pm}_{1,2}({\bf k})|^2 =
\sum_{\omega \in S} |\Psi({\bf k}, \omega^\pm_{1, 2})|^2 \, ,
\label{eq: amplitudy}
\end{equation}
where we sum over all frequencies within a range of peaks centered at $\omega^\pm_{1, 2}$. The form (\ref{eq: przyblizenie num}) is similar to the one given by the Bogoliubov transformation, Eqs. (\ref{eq: filed P}) and (\ref{eq: filed}).

In order to determine a number of quasiparticles in every mode we assume that they are related to amplitudes of the classical field Eq.(\ref{eq: amplitudy}) through the Bogoliubov transformation. Using formulas (\ref{eq: filed P}) and (\ref{eq: filed}) we get the following expression for a number of quasiparticles of the first type:
\begin{equation}
n_{1}({\bf k}) \equiv \frac{N_{1}({\bf k})}{Q} =
\frac{|\mathcal{A}^{+}_1({\bf k})|^2}{|f_{\bf k}|^2} \, ,
\end{equation}
and the second type:
\begin{equation}
n_{2}({\bf k}) \equiv \frac{N_{2}({\bf k})}{Q}=
\frac{|\mathcal{A}^{-}_2({\bf k})|^2}{|g_{\bf k}|^2} \, .
\end{equation}
According to the definitions Eqs.(\ref{eq: fk}-\ref{eq: gk}) the coefficients $f_{\bf k}$ and $g_{\bf k}$ depend on the chemical potential and the condensate amplitude at zero temperature. However, at nonzero temperatures we shall use the finite temperatures values of $\mu(T)$ and $|\phi_0(T)|^2$ obtained numerically from the spectrum of $k=0$ mode. This is a very important assumption of our approach which determines the link between number of particles and the classical field amplitudes. When the temperature is very low the assumption is quite obvious as it is based on the Bogoliubov approximation. For higher temperatures this assumption is a postulate borrowed from the nonrelativistic implementation of the classical fields method. As we will show below, this way of obtaining a number of quasiparticles leads to equipartition of energy which is typical for a thermal equilibrium of a classical system. 

Therefore, the relative energy (normalized to the total charge) accumulated in a given momentum mode is obviously: 
\begin{equation}
\varepsilon_{1}({\bf k}) = \omega_1({\bf k}) \, n_{1}({\bf k}) \, ,
\end{equation}
for the first kind of quasiparticles, and:
\begin{equation}
\varepsilon_{2}({\bf k}) = \omega_2({\bf k}) \, n_{2}({\bf k}) \, ,
\end{equation}
for the second kind.
\begin{figure}[]
\centerline{\includegraphics[width=0.38\textwidth, height=0.38\textheight, keepaspectratio]{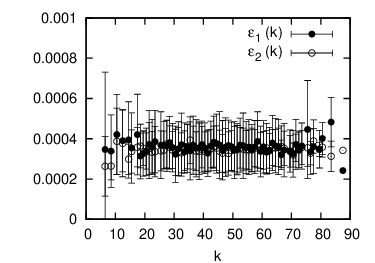}}
\caption{Equipartition relation for the first (full circles) and
for the second (empty circles) type of quasiparticles. 
Error bars refer to fluctuations of relative energy accumulated in modes with ${\bf k}=|{\bf k}|$.
The parameters are: $Q=1.25 \times 10^5\, N_{cut}$, 
$T=47.5 \, N_{cut}$, 
$m=60, \, \lambda Q =900, \, E=75$ and $\mathcal{N}^3=16^3$.}
\label{fig: equipartition}
\end{figure}
These relative energies $\varepsilon_{1}({\bf k})$ and $\varepsilon_{2}({\bf k})$ are shown in figure \ref{fig: equipartition} as a function of momentum $k$. The energies were averaged over all directions of the wave-vector ${\bf k}$ and error bars represent fluctuations of the relative energy distribution for various directions of momentum. 
The main observation is that the averaged energy per mode is constant. Moreover, it is the same for both type of quasiparticles $\varepsilon_{1}({\bf k})  \simeq \varepsilon_{2}({\bf k})$. This proves the equipartition of energy in the system -- the major property of a thermal equilibrium state of a classical system. The relative energy per mode is related to the scaled temperature $\tilde{T}$ (because we scaled the field):
\begin{equation}
\langle \varepsilon_{1} \rangle \simeq \langle \varepsilon_{2} \rangle \simeq \tilde{T} = T/Q,
\end{equation}
where $\langle \varepsilon_{1,2} \rangle$ corresponds to the energy per mode averaged over all momenta.
The scaled temperature is $\tilde{T}=3.8 \times 10^{-4}$ \cite{temp unit} for parameters of the simulation presented in figure \ref{fig: equipartition}.

Having determined the scaled temperature we can find its absolute value $T$ if the total charge of the system is known. 
To determine the charge we use the condition of macroscopic occupation of all classical modes. 
According to the previous study this can be done by requiring that the occupation of the 
highest momentum mode:
\begin{equation}
N_{cut} \equiv  N_{1}({\bf k}_{max})= Q n_{1}({\bf k}_{max}),
\end{equation}
is of the order of one $N_{cut} \simeq 1$. 
This conditions assures that the occupation of every other mode considered in the evolution ($k < k_{max}$) is evidently macroscopic. Evaluated numerically averaged relative occupations of the highest momentum mode for both type of quasiparticles
are almost the same $n_{1}({\bf k}_{max}) \simeq n_{2}({\bf k}_{max})$. 
For the case studied in this section $ n_{1,2}({\bf k}_{max}) \cong 8 \times 10^{-6}$.
Thus the field's charge is equal to 
$Q=N_{cut}/n_{1,2}({\bf k}_{max}) =1.25\times 10^5\, N_{cut}$ 
what gives the temperature $T = Q \tilde{T}=47.5 \, N_{cut}$ 
and the nonlinearity coefficient equal to $\lambda = 900/Q = 0.0072/N_{cut}$.

Notice that in the classical fields method the absolute temperature $T$ and the total charge $Q$ depend on the value of the population at the cut-off momentum. Arguments leading to the formulation of the classical fields method limit 
a value of $N_{cut}$ to the number of the order of one. In the nonrelativistic case comparison with the ideal gas 
suggests that $N_{cut} \sim 0.7$ \cite{zawitkowski, witkowska}.

\subsection{A choice of the cut-off parameter}
\label{sec: two}

The last issue we want to address here is the choice of the value of the population at the cut-off momentum $N_{cut}$. In the nonrelativistic case there is no rigorous way for choosing the cut-off parameter. Therefore, some assumptions for estimation of the value of $N_{cut}$ are in usage. Thus, the value of $N_{cut}$ is determined by assuming that the critical temperature calculated within the classical fields method is equal to those of the ideal gas \cite{zawitkowski}. This assumption seems reasonable in the relativistic case. 

\begin{figure}[t]
\centerline{\includegraphics[width=0.38\textwidth, height=0.38\textheight, keepaspectratio]{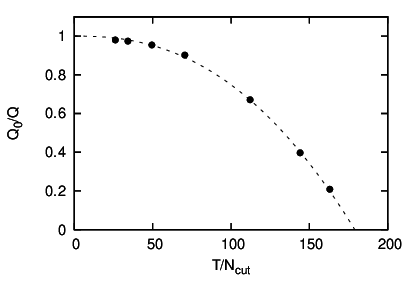}}
\caption{Charge accumulated in the zero momentum mode $Q_0 / Q$. 
Points refer to the numerical simulations while line represent the best fitted function.
Parameters of the simulations are: $m=60$, $Q\lambda=900$, $Q=1.25 \times 10^5 \, N_{cut}$, total energy and number of grid points are respectively (from the lowest temperature): $E=65$, $\mathcal{N}^3=10^3$; $E=67$, $\mathcal{N}^3=12^3$; $E=75$, $\mathcal{N}^3=16^3$; $E=96$, $\mathcal{N}^3=20^3$; $E=200$, $\mathcal{N}^3=30^3$; $E=360$, $\mathcal{N}^3=36^3$; 
$E=496$, $\mathcal{N}^3=40^3$.}
\label{fig: n0}
\end{figure}

The critical temperature can be determined from the analysis of the charge $Q_0$ assembled in the zero momentum mode. This quantity is shown in Fig. \ref{fig: n0}. Points refer to the classical fields results while lines represent the best fit. A significant fraction of the charge accumulates in the zero momentum mode below $T \sim 160 \, N_{cut}$ and tends to one at $T=0$. Let us notice that in a finite system, as considered here, there is no phase transition. The charge accumulated in the zero momentum component of the field does not vanish at any finite temperature. It rapidly decreases and at some characteristic temperature a curvature of the function $Q_0(T)$ changes sign and the charge slowly goes to zero when the temperature tends to infinity. This feature can be used to define a characteristic temperature. In the nonrelativistic case (dilute gas) this characteristic temperature becomes the critical temperature in the thermodynamic limit \cite{temperatura krytyczna}. Unfortunately the classical fields methods cannot be applied above transition temperature (no macroscopically occupied modes exist) and we cannot explore the region of slow decay of the charge accumulated in a zero momentum field. Therefore we can estimate the characteristic temperature by fitting a function to the numerical points what gives: $T_c\simeq 179 \, N_{cut}$. The critical temperature of the ideal gas is given by $T_c^0=(3\, Q/m)^{1/2}$, \cite{nieoddzialujacy RKBE}. Thus by comparing values of $T_c$ and $T_c^0$ we obtain $N_{cut} \simeq 0.2$ for the studied parameters. 

\begin{figure}[t]
\centerline{\includegraphics[width=0.38\textwidth, height=0.38\textheight, keepaspectratio]{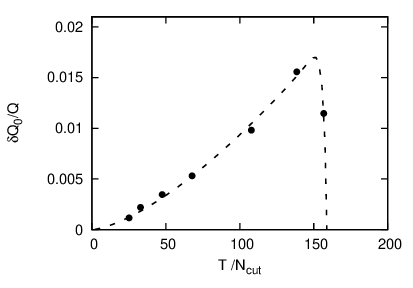}}
\caption{Fluctuations of the charge in the zero momentum mode $\delta Q_0/Q$ 
as a function of the temperature. 
Points refer to the numerical simulations while the line is plotted just to guide an eye and estimate a position of the maximum. Parameters of the simulations are the same as in figure \ref{fig: n0}.}
\label{fig: dn0}
\end{figure}

Determination of the characteristic transition temperature is not unique for the finite system as considered here. Another possibility is based on the analysis of the fluctuations $\delta Q_0 = (\langle Q_0^2 \rangle - \langle Q_0 \rangle^2)^{1/2}$ of the charge in the zero momentum mode. In Fig. \ref{fig: dn0} we show fluctuations of the charge $\delta Q_0$ versus temperature $T$. The fluctuations increase monotonically with the temperature, reach a maximum, and then rapidly fall to a very small values. The characteristic transition temperature can be defined as the temperature corresponding to the maximum of fluctuations of the charge $Q_0$. According to this definition the characteristic transition temperature is $T_0 \sim 150\, N_{cut}$. The temperature $T_0$ is smaller than $T_c$ (thus the value of $N_{cut}$ cannot be treated as the universal number). This difference results from finite size of the system. We expect that in the thermodynamic limit both these temperatures become equal as it is in the nonrelativistic case \cite{temperatura krytyczna}.

\section{Quasiparticles at nonzero temperature}
\label{sec: dependence}

\begin{figure}[]
\centerline{\includegraphics[width=0.4\textwidth, height=0.4\textheight, keepaspectratio]{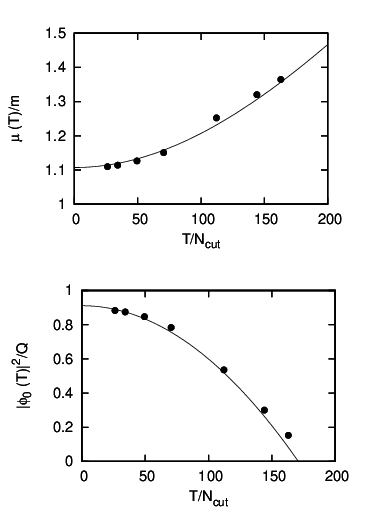}}
\caption{$(a)$ Chemical potential $\mu/m$ and $(b)$ condensate amplitude
$|\phi_0|^2/Q$. 
Points refer to the numerical simulations.
Solid lines are Eq.(\ref{eq: mu in T}) for the chemical potential,
and Eq.(\ref{eq: phi^2 in T}) for the condensate field intensity.
Parameters of the simulations are the same as in figure \ref{fig: n0}.}
\label{fig: mu}
\end{figure}

Interactions change a character of elementary excitations in the system. The first quasiparticles in the small momentum limit are phonons and therefore their nature is different from relativistic massive particles. For the quasiparticles of the second kind the role of interactions is to effectively modify a bare mass. Within the classical fields method we can investigate how both the interactions and the temperature affect the physical proprties of the system. To this aim, we will use Bogoliubov-Popov like relations: Bogoliubov formulas determined in the section III with temperature dependent chemical potential $\mu(T)$ and condensate amplitude $|\phi_0(T)|^2$. This two last quantities are crucial for the whole analysis. In Fig. \ref{fig: mu} we show $\mu(T)$ and $|\phi_0(T)|^2$ calculated numerically (points) and we compare them to the results of other authors \cite{{kapusta},{Andersen}} (lines). The value of the chemical potential $\mu$ increases with the temperature as it is predicted by \cite{kapusta}:
\begin{equation}
\mu(T) = \sqrt{\mu^2 + \frac{\lambda}{3} \left(  \frac{T}{m }\right)^2}\, ,
\label{eq: mu in T}
\end{equation}
where the zero temperature chemical potential $\mu$ results from the Bogoliubov approximation (mass $m$ in denominator results from our definition of the charge (\ref{eq: charge})). On the other hand condensate amplitude $|\phi_0(T)|^2$ decreases with the temperature in the way given by the formula \cite{Andersen}:
\begin{equation}
\frac{|\phi_0(T)|^2}{Q}=\frac{|\phi_0|^2}{Q} - \left(\frac{T}{T_c} \right)^2 \, ,
\label{eq: phi^2 in T}
\end{equation}
where $|\phi_0|^2$ is the Bogoliubov approach result. We observed quantitative agreement between the classical fields method and predictions of \cite{kapusta} and \cite{Andersen}. Having determined $\mu(T)$ and $|\phi_0(T)|^2$ in the whole range of interesting temperatures we can look for thermal properties of the system. We concentrate on four quantities: the excitation energy spectra, decay rates, mixing parameters and charges of quasiparticles.

\begin{figure}[]
\centerline{\includegraphics[width=0.38\textwidth, height=0.38\textheight, keepaspectratio]{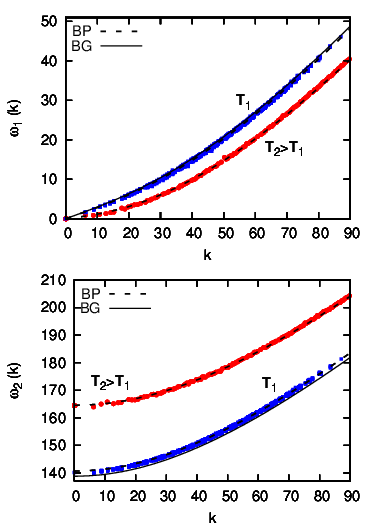}}
\caption{(Color online) Excitation energy $\omega_1({\bf k})$
and $\omega_2({\bf k})$ as a function of momentum $k=|{\bf k}|$.
Blue points refer to the numerical simulations for 
$E_1=75$, $T_1=47.5 N_{cut}$, and the number of spatial grid points 
$\mathcal{N}^3=16^3$;
red points refer to the numerical results with
$E_2=500$, $T_2=163.0 N_{cut}$, and the number of grids points 
$\mathcal{N}^3=40^3$.
Solid line corresponds to the Bogoliubov approximation (marked by $BG$),
and dashed line to the Bogoliubov-Popov like formulas.}
\label{fig: w1w2}
\end{figure}
\begin{figure}[]
\centerline{\includegraphics[width=0.4\textwidth, height=0.4\textheight, keepaspectratio]{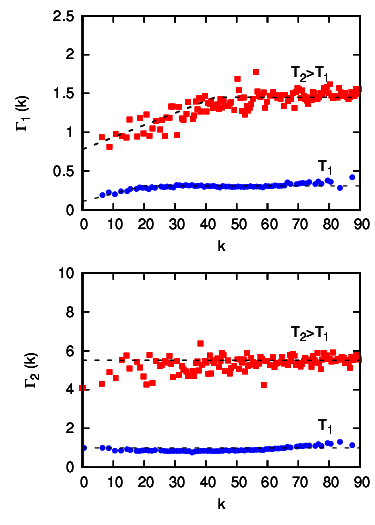}}
\caption{(Color online) Decay rates as a function of momentum: $\Gamma_1(k)$ for the first type of quasiparticles and 
$\Gamma_2(k)$ for the second quasiparticles. Lines are added to guide the eye.
Numerical simulations results for two different temperatures:
$T_1=47.5 N_{cut}$, $E_1=75$ and $\mathcal{N}^3=16^3$ and $T_2=163.0 N_{cut}$, $E_2=500$ and $\mathcal{N}^3=40^3$.}
\label{fig: damping}
\end{figure}

First, in Fig. \ref{fig: w1w2} we show excitation spectra $\omega_1(\bf k)$ and $\omega_2(\bf k)$ for two different temperatures. Points refer to classical fields results and lines are: Bogoliubov (solid) and Bogoliubov-Popov like (dashed) formulas. For the lower temperature, $T_1$, all results are very close to each other. The phonon branch in the first type of quasiparticles excitation spectrum is clearly visible. For the higher temperature, $T_2$, the phonon branch practically disappears because the sound velocity significantly decreases.  These features are well known in the nonrelativistic limit \cite{cfm1a}. The excitation spectrum of the second type of quasiparticles does not have any visible linear part at long wavelength. The finite lifetime of excitations are determined by decay rates of quasiparticles. They are related to the width of the peaks in the excitation spectrum (see Fig. \ref{fig: spectrum of amplitudes1}). The finite width indicates a finite lifetime of quasiparticles and signifies a weak residual interactions between quasiparticles. The peak width can be estimated by:
\begin{equation}
\Gamma_{1,\, 2}({\bf k}) = 0.5 \sqrt{\langle \omega_{1,\, 2}^2({\bf k}) \rangle - \langle \omega_{1,\, 2}({\bf k}) \rangle^2}
\end{equation} 
where $\langle \cdot \rangle$ represents averaging over frequencies within a peak area, compare Eq.(\ref{eq: omega cfs}). In Fig. \ref{fig: damping} we show decay rates of the first, $\Gamma_1$, and the second, $\Gamma_2$, type of quasiparticles as functions of $k$. One can see that in the case of the first type of quasiparticles decay rates are increasing in the small momenta region where the dispersion relation is linear. The decay rates weakly depend on momentum for large $k$. This weak dependence is characteristic for the second type of quasiparticles in the whole momenta range. Moreover, the decay rates of the first type of quasiparticles are much smaller than the decay rates of the quasiparticles of the second type. In addition, the decay rates of both quasiparticles types increase with the temperature.

\begin{figure}[]
\centerline{\includegraphics[width=0.38\textwidth, height=0.38\textheight, keepaspectratio]{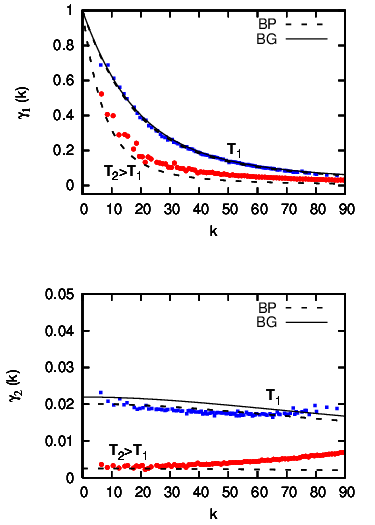}}
\caption{(Color online) 
Coefficients $\gamma_1({\bf k})$ and $\gamma_2({\bf k})$ of the Bogoliubov transformation
(\ref{eq: gamma1}) and (\ref{eq: gamma2}) as a function of $k=|{\bf k}|$.
Points refer to numerical simulations results for two different temperatures:
$T_1=47.5 N_{cut}$, $E_1=75$ and $\mathcal{N}^3=16^3$ (blue); and $T_2=163.0 N_{cut}$, $E_2=500$ and $\mathcal{N}^3=40^3$ (red).
Solid line ($BG$) refers to the Bogoliubov approximation and dashed line
to the Bogoliubov-Popov like formulas (BP).}
\label{fig: g1g2}
\end{figure}
The relativistic Bogoliubov transformation diagonalizing approximated Hamiltonian `mixes' particle's creation and annihilation operators: annihilation operator of a quasiparticle is a superposition of annihilation operator of a bare particle with momentum ${\bf k}$ and creation operator of a bare particle with momentum $-\bf k$. The amount of this mixing is given by the parameter $\gamma_1(\bf k)$ for the first type of quasiparticles, and $\gamma_2(\bf k)$ for the second type. The coefficients of the relativistic Bogoliubov transformation can be calculated numerically in the following way:
\begin{eqnarray}
\gamma_1({\bf k}) &=& \frac{|\mathcal{A}^{-}_1({\bf k})|}{|\mathcal{A}^{+}_1({\bf k})|} \, , \\
\gamma_2({\bf k}) &=& \frac{|\mathcal{A}^{+}_2({\bf k})|}{|\mathcal{A}^{-}_2({\bf k})|} \, ,
\end{eqnarray}
where $\mathcal{A}^{\pm}_{1 \, 2}$ are amplitudes defined in (\ref{eq: przyblizenie num}).
On the other hand they can be calculated from the Bogoliubov-Popov formulas (23), (24).
In Fig. \ref{fig: g1g2} we show $\gamma_1$ and $\gamma_2$. Points refer to results averaged over angles of the momentum ${\bf k}$ for two different temperatures: $T_1=47.5 N_{cut}$ marked by blue, and $T_2=163.0 N_{cut}$ marked by red. The solid line is the result of the Bogoliubov procedure while dashing line corresponds to the Bogoliubov-Popov like relations. The results show that in the linear dispersion regime (phonon spectrum) the value of coefficient $\gamma_1$ is close to one. The excitations have a quasiparticle character. In a region of large momenta $\gamma_1$ is small and quasiparticles are equivalent to bare particles. For the excitations of the second kind $\gamma_2$ is very small for all momenta and decreases with increasing temperature. This indicates that distinction between the second quasiparticle types and bare antiparticles is practically meaningless for the values of parameters studied here.

\begin{figure}[]
\centerline{\includegraphics[width=0.38 \textwidth, height=0.38 \textheight, keepaspectratio]{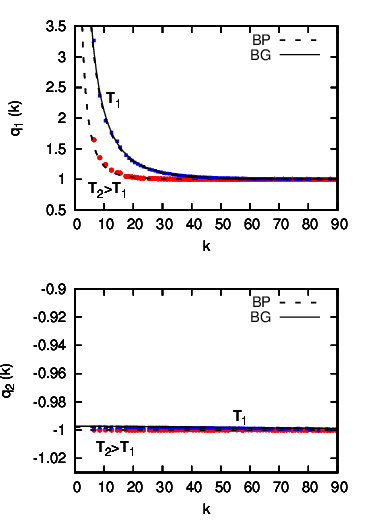}}
\caption{(Color online) Effective charges of the first quasiparticle types $q_1(k)$ and 
the second quasiparticle types $q_2(k)$ for temperatures: $1)$ $T_1=47.5 N_{cut}$ (blue points) 
and $2)$ $T_2=163.0 N_{cut}$ (red points).  Points refer to equations (\ref{eq: qP}), (\ref{eq: qA}) with values of $\omega_1$, $\omega_2$, $\mu$ and $|\phi_0|^2$ determined numerically,
dashed lines correspond to the Bogoliubov-Popov like formulas, and solid line refers to the Bogoliubov result.}
\label{fig: ladunek}
\end{figure}
Finally, let us look for effective charge $q_1({\bf k})$ and $q_2({\bf k})$ of quasiparticles. 
In Fig. \ref{fig: ladunek} we show these effective charges as functions of momentum $k$. All quasiparticles of the second type have practically the same charge $q_1(k) \simeq -1$ independently on the value of the temperature or momentum. The same charge is carried by an antiparticle in the noninteracting ($\lambda=0$) case \cite{Bjorken}. On the other hand the charge $q_1(k)$ carried by a single quasiparticle of the first type strongly depends on its momentum. It is greater than one $q_1>1$ for the phonon-like quasiparticles while it is equal to one for the quasiparticles with large momentum. Highly exited quasiparticles of the first type have the same charge as particles in the noninteracting case. Our simulations show that the quasiparticle charges decreases with increasing temperature.

\section{Summary}
\label{sec: con}

In this paper we presented the classical fields method for description of a relativistic interacting Bose gas at thermal equilibrium. First, we studied the system at zero temperature using Bogoliubov method. We identified two quasiparticles types. Then we formulated the classical fields method. We postulated that the classical field has a structure of the Bogoliubov transformation. This allows us to relate $k$-momentum amplitude to the occupation number of the $k$-mode. By showing equipartition of energy and the structure of the mixing parameters $\gamma_{1, \, 2}$ we verified positively our postulate. In addition, we discussed some technical details important for the correct implementation of the method. The classical fields method is an approximate one, similarly to the others, and allows for qualitative but nonperturbative study of thermal properties of the system. In the case of the relativistic interacting Bose gas, the classical fields simulations show that the system forms a weakly interacting gas of two quasiparticles types not only at zero but in the whole range of temperatures. 

{\bf Acknowledgment}
We thank prof. Jan Mostowski for stimulating discussion. E.W. and M.G. acknowledge support by the Polish Government research funds for 2006-2009. P.Z. thanks Polish Government research funds for 2007-2010.


\begin{thebibliography}{99}

\bibitem{exact}
see for instance: M. Holland, K. Burnett, C.W. Gardiner, J.I. Cirac and P. Zoller,
Phys. Rev. A, 54, R1757, (1996);
C.W. Gardiner, P. Zoller, Phys. Rev. A, 552902, (1997);
P.D. Drummond, J.F. Corney, Phys. Rev. A, 60 R2661, (1999);
I. Carusotto, Y. Castin J. Phys. B: At.Mol.Opt. Phys. 34 4583 (2001).

\bibitem{cfm1}
K. G\'oral, M. Gajda, K. Rz\k{a}\.zewski, \emph{Opt. Express}, \textbf{8}, 92 (2001).

\bibitem{cfm1a}
M. Brewczyk, P. Borowski, M. Gajda, K. Rz\k{a}\.zewski, \emph{J. Phys. B}, \textbf{37}, 2725 (2004).

\bibitem{cfm2}
A. Sinatra, C. Lobo, Y. Castin, \emph{Phys. Rev. Lett.}, \textbf{87}, 210404 (2001).

\bibitem{wigner} 
A. Sinatra, C. Lobo, Y. Castin, \emph{J. Phys. B: At. Mol. Opt. Phys.}, \textbf{35}, 3599-3631 (2002).

\bibitem{cfm3}
M. J. Davis, S. A. Morgan, and K. Burnett, \emph{Phys. Rev. Lett.}, {\bf 87}, 160402 (2001);

\bibitem{gardiner}
Gardiner CW, Anglin JR, Fudge TIA, \emph{J. Phys. B: At. Mol. Opt. Phys.}, {\bf 35} 1555-1582 (2002)

\bibitem{zaremba}
Zaremba E, Nikuni T, Griffin A, \emph{Journal of low temperature physics}, {\bf 116} 277-345 (1999)

\bibitem{review}
M. Brewczyk, M. Gajda, K. Rz\k{a}\.zewski, 
\emph{J. Phys. B: At. Mol. Opt. Phys.}, \textbf{40}, No 2 R1-R37 (2007).


\bibitem{Davis}
M. J. Davis and S. A. Morgan, Phys. Rev. A  \textbf{68} 053615 (2003).

\bibitem{schmidt}
H. Schmidt, K. G\'oral, F. Floegel, M. Gajda, K. Rz\k{a}\.zewski
\emph{J.Opt.B: Quantum Semiclass. Opt.} {\bf 5} S96 (2003).

\bibitem{Kadio}
D. Kadio, M. Gajda, K. Rz\k{a}\.zewski, \emph{Phys. Rev. A} {\bf 72} 013607 (2005).

\bibitem{lobo}
Carlos Lobo, Alice Sinatra, and Yvan Castin,
\emph{Phys. Rev. Lett.} {\bf 92}, 020403 (2004).

\bibitem{Gawryluk}
K. Gawryluk, M. Brewczyk and K. Rz\k{a}\.zewski
\emph{J. Phys. B: At. Mol. Opt. Phys.} {\bf 39} L225-L231 (2006).

\bibitem{Zawitkowski-sup} 
\L{}. Zawitkowski, M. Gajda, and K. Rz\k{a}\.zewski
\emph{Phys. Rev. A} {\bf 74}, 043601 (2006).

\bibitem{spinor}
K. Gawryluk, M. Brewczyk, M. Gajda, and K. Rz\k{a}\.zewski
\emph{Phys. Rev. A} {\bf 76}, 013616 (2007).

\bibitem{KBT}
T. P. Simula and P. B. Blakie, Phys. Rev. Lett. 96, 020404 (2006).

\bibitem{Sinatra & Witkowska}
A. Sinatra, Y. Castin, E. Witkowska, \emph{Phys. Rev. A} {\bf 75}, 0033616 (2007);
A. Sinatra, Y. Castin, http://hal.archives-ouvertes.fr/hal-00310488/fr .

\bibitem{Gorecka} A. G\'orecka, M. Gajda, subbmited to \emph{Phys. Rev. A}.

\bibitem{Bjorken}
J. D. Bjorken, S. D. Drell, \emph{Relativistic Quantum Mechanics}, McGraw-Hill, Inc., (1964).

\bibitem{BjorkenGreiner}
W. Greiner, \emph{Relativistic quantum mechanics}, 3th Edition, Springer-Verlag Berlin Heidelberg, (2000).

\bibitem{noninteracting1}
H. Feshbach and F. Villars, \emph{Rev. Mod. Phys.}, {\bf{30}}, no 1, 24 (1958).

\bibitem{noninteracting2}
P. T. Landsberg and J. Dunning-Davies, \emph{Phys. Rev. A}, \textbf{138} 1049 (1965).

\bibitem{noninteracting3}
R. Beckmann, F. Karsch, D. E. Miller, \emph{Phys. Rev. Lett.}, \text{43} 1277 (1979).

\bibitem{nieoddzialujacy RKBE}
H.E. Haber i H.A. Weldon, \emph{Phys. Rev. Lett.} {\bf 46}, 1497 (1981).

\bibitem{noninteracting5}
L. Salasnich, \emph{Nuovo Cim.}, 117B 637 (2002).

\bibitem{noninteracting6}
L. Burakovsky, L.P. Horvitz, W.C. Schieve, \emph{Phys. Rev. D}, {\bf 54} 6 (1996);
K. Shiokawa, B.L. Hu, \emph{Phys. Rev. D}, {\bf 60}, 105016 (1999);
V.V. Begun, M.I. Gorenstein, \emph{Phys. Rev. C}, {\bf 77}, 064903 (2008)

\bibitem{Boisseau}
B. Boisseau, \emph{J. Phys. A: Math. Gen.}, \textbf{37}, 7923 (2004).

\bibitem{interacting}
G. Su, J. Chen, L. Chen, \emph{J. Phys. A: Math. Gen.}, \textbf{39}, 4935 (2006);
M. Grether,M. de Llano and G.A. Baker, \emph{Phys. Rev. Lett.}, {\bf {99}}, 200406 (2007);
G. Su, Sh. Cai, J. Chen, \emph{J. Phys. A: Math. Gen.}, \textbf{41}, 045007 (2008).

\bibitem{Bernstein}
J. Bernstein, S. Dodelson, \emph{Phys. Rev. Lett.}, \textbf{66}, 683 (1991).

\bibitem{Haber}
H.E. Haber, H.A. Weldon, \emph{Phys. Rev. D}, \textbf{25}, 502 (1982).

\bibitem{kapusta}
J.I. Kapusta, \emph{Phys. Rev. D}, {\bf 24} 426 (1981).

\bibitem{Andersen}
J. O. Andersen, \emph{Phys. Rev. D}, {\bf 75} 065011 (2007).


\bibitem{other} 
Hua Shi, Allan Griffin, \emph{Physics Reports}, {\bf 304}, 1-87 (1998);
S.T. Beliaev, {\it JETP-USSR}, {\bf 7} no.2, 299-307 (1958);
F. Mohling  and M. Morita, \emph{Phys. Rev.}, {\bf 120}, 681 (1960); 
P. O. Fedichev, G. V. Shlyapnikov, and J. T. Walraven, \emph{Phys. Rev. Lett.}, {\bf 80}, 2269 (1998);
P. O. Fedichev and G. V. Shlyapnikov, \emph{Phys. Rev. A}, {\bf 58}, 3146 - 3158 (1998).

\bibitem{Equilibrium NLKGE}
see for instance:
A. Arrizabalaga, J. Smit, and A. Tranberg \emph{Phys. Rev. D} {\bf 72} 025014 (2005);
A. Arrizabalaga, \emph{The European Physical Journal A} {\bf 29} 101 (2006).

\bibitem{1+1}
G. Aarts and J. Smit,
\emph{Phys. Rev. D} {\bf 61} 025002 (1999);
D. Boyanovsky, C. Destri and H.J. de Vega \emph{Phys. Rev. D}
{\bf 69} 045003 (2004).

\bibitem{witkowska}
E. Witkowska, M. Gajda, J. Mostowski, \emph{J. Phys. B: At. Mol. Opt. Phys.}, \textbf{40} 1465 (2007).

\bibitem{Bogoliubov}
N.N. Bogoliubov, \emph{J. Phys.}, \textbf{11}, 23 (1947).

\bibitem{units} 
In this paper the following units are used:
$1)$ length $\tilde{{\bf r}}=V^{1/3} {\bf r}$, 
$2)$ time $\tilde{t}=t V^{1/3}/c$, $3)$ field $\tilde{\Psi}=\Psi/\sqrt{V}$, 
$4)$ mass $\tilde{m} = \hbar m/c V^{1/3}$, $5)$ $\tilde{\Lambda}=\lambda V^{1/3}$,
and energy $\tilde{H}=H \hbar c/V^{1/3}$, where:
${\bf r}, \, t, \, \Psi, \, m, \, \lambda$ and $H$ are dimensionless.

\bibitem{quantum depletion value}
The value of a quantum depletion is $\mathcal{C}/Q=0.00025$
for parameters studied in the numerical part:
$m=60$, $\lambda Q =900$ and ${\bf k}_{max}=2 \pi (8,8,8)$.

\bibitem{zawitkowski}
\L{}. Zawitkowski, M. Brewczyk, M. Gajda, K. Rz\k{a}\.zewski, \emph{Phys. Rev. A}, \textbf{70}, 033614 (2004).

\bibitem{Green comment}
Analysis of Green function singularities allows to determine excitation energies and number of quasiparticles \cite{fetter} in the many-body physics. In our case the Bogoliubov transformation gives yet another possibility to identified numerically quasiparticles: its energies excitation and their number. This is done in the section \ref{sec: simulations}.

\bibitem{fetter} A. L. Fetter, J. D. Walecka, \emph{Quantum theory of many-particle systems}, McGraw-Hill Inc. (1971).

\bibitem{BP} 
K. Burnett, M. Edwards, C. W. Clark, \emph{Physics Today} 37 (1999).

\bibitem{temp unit}
The temperature unit is: $\hbar c/V^{1/3} k_{B}$, where $k_B$ is Boltzmann constant.

\bibitem{temperatura krytyczna}
Z. Idziaszek, K. Rz\k{a}\.zewski, \emph{Phys. Rev. A}, {\bf 68}, 035604 (2003).

\bibitem{Alber} G. Alber, \emph{Phys. Rev. A}, {\bf 63}, 023613 (2001).



\end{thebibliography}
\end{document}